\documentstyle[twoside,natbib,epsf]{article}

\input{ibvs2.sty}

\def\cite#1{\citealt{#1}}
\def\ibvs{Inf. Bull. Var. Stars}
\def\aap{A\&A}
\def\aaps{A\&AS}

\def\apjl{ApJ}
\def\apjs{ApJS}

\def\mnras{MNRAS}
\def\pasj{PASJ}
\def\pasp{PASP}
\def\PVSS{Publ. of Var. Star Sec., RASNZ}

\def\inpress{in press}
\def\astroph#1{ (astro-ph/#1)}

\begin{document}

\IBVShead{xxxx}{xx July 2002}

\IBVStitle{Unusually Increased Activity of GZ C\lowercase{nc}}

\IBVSauth{Kato,~Taichi$^1$, Dubovsky,~Pavol~A.$^2$, Stubbings,~Rod$^3$, Simonsen,~Mike$^4$
}
\vskip 5mm

\IBVSinst{Dept. of Astronomy, Kyoto University, Kyoto 606-8502, Japan, \\
          e-mail: tkato@kusastro.kyoto-u.ac.jp}

\IBVSinst{MEDUZA group, Vedecko-kulturne centrum na Orave,
          027 42 Podbiel 194, Slovakia
          e-mail: vkco@isternet.sk}

\IBVSinst{19 Greenland Drive, Drouin 3818, Victoria, Australia,
          e-mail: stubbo@sympac.com.au}

\IBVSinst{46394 Roanne Drive Macomb, MI USA 48044
          e-mail: mikesimonsen@mindspring.com}

\IBVSobj{GZ Cnc}
\IBVStyp{UG}
\IBVSkey{dwarf nova, photometry, activity cycle}

\begintext

   GZ Cnc was discovered as a variable star by Takamizawa (Tmz~V34).
Subsequent observations revealed that this object is a dwarf nova
which is identified with a ROSAT source \citep{kat01gzcnc}.
The object was independently confirmed to be a cataclysmic variable
in the course of optical identifications of ROSAT bright sources
\citep{bad98RASSID}.  \citet{jia00RASSCV} reported an optical spectrum
which showed strong Balmer and He\textsc{I} emission lines.  Although
\citet{jia00RASSCV} did not explicitly mentioned, He\textsc{II} emission
lines were detected stronger than in typical dwarf novae
\citep{wil83CVspec1}.

   \citet{kat01gzcnc} obtained time-resolved CCD photometry of the 2000
February long outburst.  The long duration of the outburst and the slow
rising rate suggested that GZ Cnc is a good candidate for a long-period
dwarf nova.  As reported in \citet{kat01gzcnc}, the recorded outbursts
up to 2001 were relatively rare.  Although there were unavoidable
seasonal observational gaps, only three outbursts were recorded between
1999 and 2001.

   In 2002 March -- May, we noticed a dramatic increase of the outburst
frequency.  Figure 1 shows the light curve of the 2002 season, mainly
drawn from visual observations by the authors.  Some additional
observations (visual and CCD) reported to the VSNET Collaboration
(http://www.kusastro.kyoto-u.ac.jp/vsnet/) have been incorporated.
All observers used $V$-band calibrated comparison stars.  The uncertainties
of the observations are 0.2--0.3 mag, which will not affect the following
discussion.

\IBVSfig{9cm}{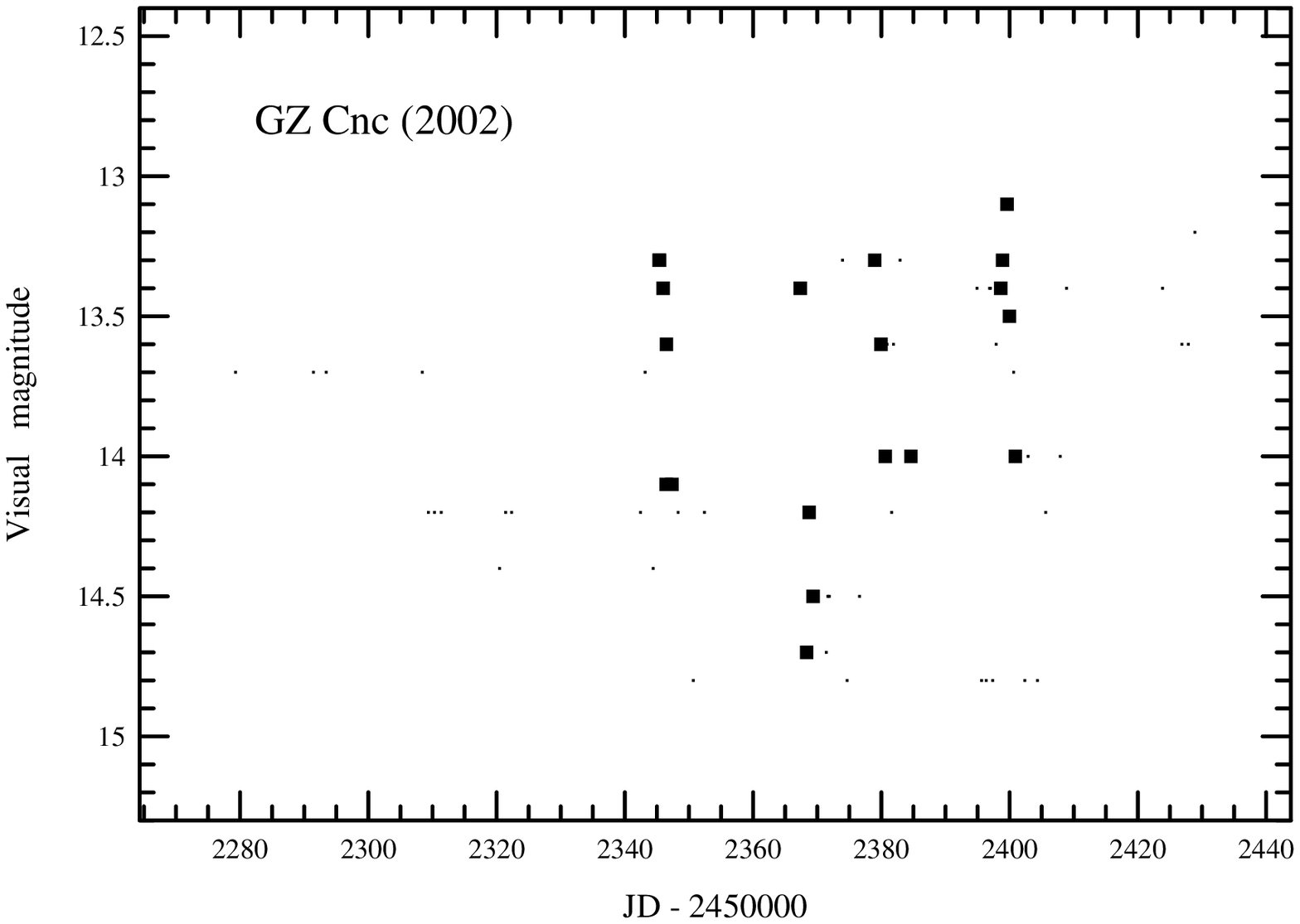}{
    Light curve of GZ Cnc in the 2002 season.  Large and small dots
represent positive and negative (upper limit) observations, respectively.
Note the high frequency of outbursts (large dots).
}

   Table 1 lists the known outbursts of GZ Cnc since the discovery by
Takamizawa.  The shortest interval outbursts in the 2002 unusually active
season was only 11 d, and the other two intervals were 21--22 d.
As shown in Figure 1, the durations of the outbursts in 2002 were
very short in contrast to the long outbursts in 2000 February
\citep{kat01gzcnc}.  Although such bimodal activity may suggest
an outburst activity seen in SU UMa-type dwarf novae
\citep{vog80suumastars,war85suuma}, the apparent lack of superhumps
during the long outburst seems to exclude the possibility of an SU UMa-type
dwarf nova \citep{kat01gzcnc}.

   Alternately, the present behavior in some aspects resembles a
``clustering" of outbursts observed in some intermediate polars (IPs)
[EX Hya: \citet{bat86exhyaoutburst,hel89exhya};
TV Col: Uemura et al. in preparation],
whose interpretation is still in debate \citep{hel00exhyaoutburst}.
Although no clear coherent pulses were detected during the 2000 February
outburst \citep{kat01gzcnc}, the presence of He\textsc{II} emission lines
and the relatively hard X-ray spectrum \citep{bad98RASSID} would encourage
a more extensive search for the IP-type coherent signal in quiescence
and outburst (e.g. \cite{kem02htcam,ish02htcam}).

\begin{table}
\begin{center}
Table 1. Outbursts of GZ Cnc. \\
\vspace{10pt}
\begin{tabular}{ccccc}
\hline
\multicolumn{3}{c}{Date} & JD-2400000 & Max \\
\hline
1994 & November & 30 & 49687 & 13.1$^a$ \\
2000 & February & 3  & 51578 & 13.7$^b$ \\
2000 & December & 29 & 51908 & 13.1 \\
2002 & March    & 11 & 52345 & 13.3 \\
2002 & April    & 2  & 52367 & 13.4 \\
2002 & April    & 14 & 52378 & 13.3 \\
2002 & May      & 4  & 52399 & 13.1 \\
\hline
 \multicolumn{5}{l}{$^a$ Discovery observation by Takamizawa.} \\
 \multicolumn{5}{l}{$^b$ Long outburst reported in \citet{kat01gzcnc}.} \\
\end{tabular}
\end{center}
\end{table}

\IBVSfig{9cm}{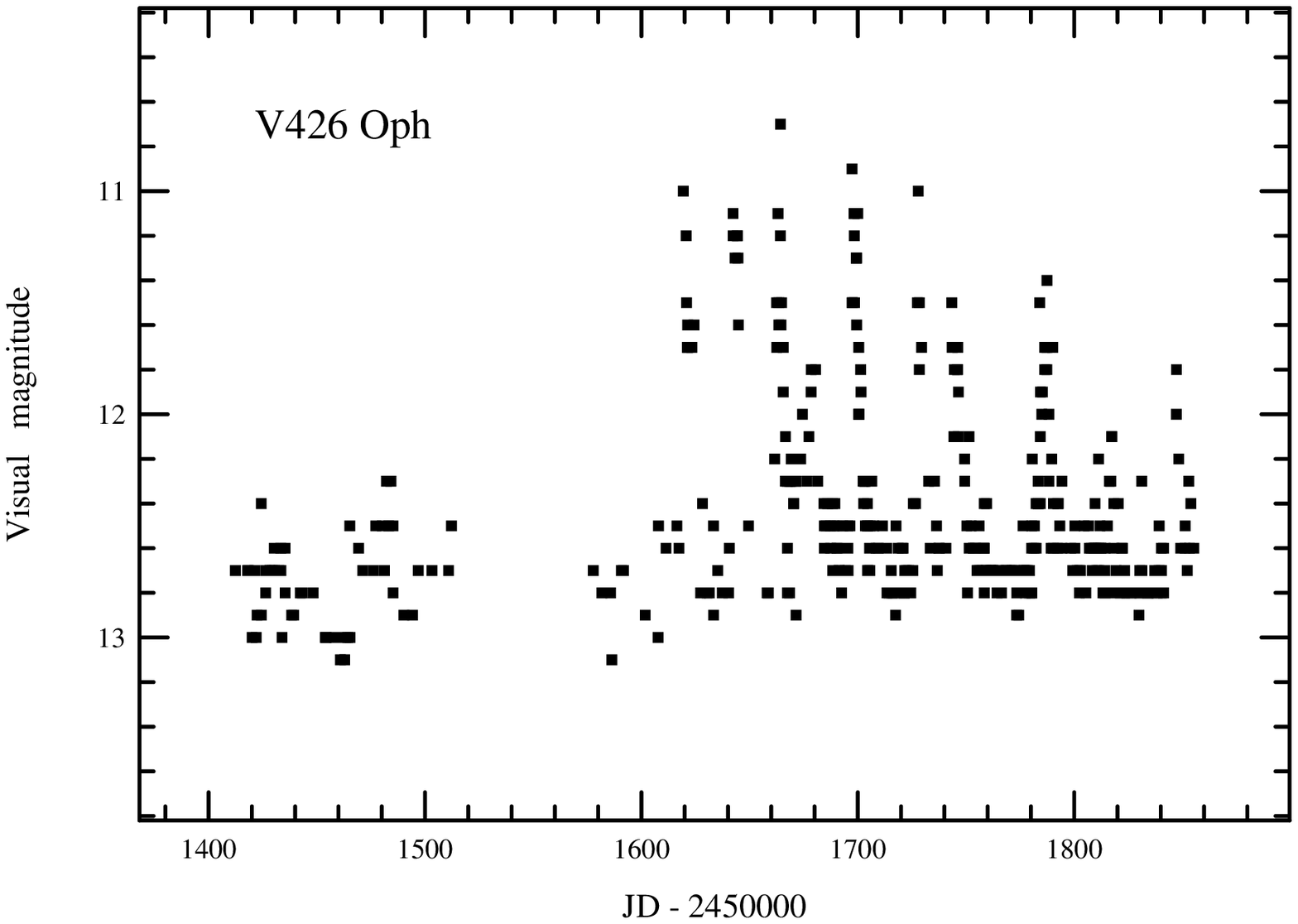}{
   An state of increased outburst activity in V426 Oph (1999--2000).
The data are from reports to VSNET.
}

   The present activity can also be comparable to V426 Oph, another
dwarf nova which is known to show occasionally increased activities
(Figure 2; see also \cite{wen90v426oph}).
V426 Oph has been also suggested to be an IP
\citep{szk86v426ophEXOSAT}, but this possibility is recently questioned
\citep{hel90v426oph}.  The relatively hard X-ray spectrum of V426 Oph
\citep{ver97ROSAT} is also suggestive of an analogy between GZ Cnc and
V426 Oph, which may comprise a new class of cataclysmic variables with
prominent occasional increases of outburst activities.  These activities
may be a result of the weak presence magnetic fields, although the
evidence of the magnetic nature (at least in V426 Oph) is still
tantalizing.  Further research to elucidate the relation between these
unusual cataclysmic variables and IPs is encouraged.

\vskip 3mm

We are grateful to VSNET observers who have reported observations of
GZ Cnc and V426 Oph.
This work is partly supported by a grant-in aid (13640239)
from the Japanese Ministry of Education, Culture, Sports,
Science and Technology.

\end{document}